# An Elemental Ethics for Artificial Intelligence: Water as Resistance Within AI's Value Chain

Dr Sebastián Lehuedé[1][2]

sebastian.lehuede@kcl.ac.uk

ABSTRACT

*Research and activism have increasingly denounced the problematic environmental record of the infrastructure and value chain underpinning Artificial Intelligence (AI). Water-intensive data centres, polluting mineral extraction and e-waste dumping are incontrovertibly part of AI's footprint. In this article, I turn to areas affected by AI-fuelled environmental harm and identify an ethics of resistance emerging from local activists, which I term 'elemental ethics'. Elemental ethics interrogates the AI value chain's problematic relationship with the elements that make up the world, critiques the undermining of local and ancestral approaches to nature and reveals the vital and quotidian harms engendered by so-called intelligent systems. While this ethics is emerging from grassroots and Indigenous groups, it echoes recent calls from environmental philosophy to reconnect with the environment via the elements. In empirical terms, this article looks at groups in Chile resisting a Google data centre project in Santiago and lithium extraction (used for rechargeable batteries) in Lickan Antay Indigenous territory, Atacama Desert. As I show, elemental ethics can complement top-down, utilitarian and quantitative approaches to AI ethics and sustainable AI as well as interrogate whose lived experience and well-being counts in debates on AI extinction.*

---

[1] Department of Digital Humanities, King's College London

[2] Centre of Governance and Human Rights, University of Cambridge

# 1. Introduction

Tuesday, 2nd March of 2022. I join the Water Day Protest in Santiago, Chile's capital. A popular slogan at the rally is '*no es sequía, es saqueo*' [this isn't a drought, this is a robbery]. Protestors range from local communities to Indigenous people and workers' unions. The march finishes in the offices of the Constituent Assembly, where topics such as the 'rights of nature' are debated by democratically-elected representatives.

This protest, which I attended as part of my fieldwork on digital rights, prompted the questions over AI, ethics and the environment that I explore in this article. But how do World Water Day, Chile and AI connect in the first place? Part of the answer lies on the group that invited me to join the rally: MOSACAT[3]. This organisation, whose members I had met a few weeks before, had formed an opposition to the construction of a water-intensive Google data centre in the working-class area of Cerrillos in Santiago. A month after this protest, I also engaged with the Council of Atacameno Peoples ('the Council' hereafter), which reunites eighteen Lickan Antay communities resisting water-intensive lithium extraction in the Atacama Desert[4].

Both data centres and minerals such as lithium are key components of digital technologies' value chains. Their construction and extraction have increased with the current wave of AI technologies. Firstly, data centres are the buildings that host the computers processing the vast amounts of data required to train AI systems and run AI applications. Such processing generates significant heat, with water circulation and electric air conditioning employed to cool off these servers. The construction of data centres is being advanced by transnational technology companies such as Google, Huawei and Amazon as a means to deal with the technical challenges posed by trends such as AI (the more data is employed to identify hidden patterns, the more energy-intensive the process becomes).

Secondly, lithium is a mineral crucial for building the rechargeable batteries that power wireless devices. Such devices, which range from mobile phones to smart assistants, are key facilitators of AI as they make it possible to generate the data required to train AI algorithms and to enable AI-powered applications. Electric vehicles, which currently drive most of the demand for lithium (Jerez et al., 2011), have been deemed as 'green technologies';

---

[3] *Movimiento Socioambiental Comunitario por el Agua y el Territorio* (Socio-Environmental Communitarian Movement for Water and Territory).

[4] The Lickan Antay people are also known as 'Atacameno' (Atacameño in Spanish) due to a misnaming by the Spanish invaders.

paradoxically, the extraction of lithium in the Atacama Desert, one of the driest areas of the world, employs vast amounts of water.

This article is part of a broader turn in the study of AI, data, algorithmic and digital technologies bringing to the fore the social and environmental issues taking place at the level of infrastructure and the value chain (e.g., Taffel, 2015; Parikka, 2015; Hogan, 2015; Gabris, 2011). In relation to this literature, I seek to make a contribution by taking seriously the needs and visions of the communities negotiating with AI's value chain in order to propose an alternative ethics for the design and governance of so-called 'intelligent' systems. In so doing, I am avoiding portraying these communities as mere 'victims' but am approaching them as holders of valuable and legitimate knowledge in times of AI and climate crises. This move is crucial since 'solutions' to the environmental issues brought about by AI envisioned by a narrow set of experts are deemed to fail as long as they do not speak to and incorporate the experience of affected communities. A particular challenge I faced in this regard was that terms such as 'AI' or 'digital technologies' were not central in the activism of MOSACAT or the Council. However, I show here that these communities' vigorous protection of water (as both data centres and lithium extraction require vast amounts of it) could inspire a sustainable, grounded and bottom-up ethical approach to AI.

Conceptually speaking, in this article I contend that elemental thought can bridge debates on AI ethics with the activism undertaken by groups such as MOSACAT and the Council. In recent years, environmental and science and technology studies scholars have drawn attention to the generative possibilities of rethinking the relationship between humans and the environment via the latter's constituent parts, namely through elements such as water, fire, air and earth (Macauley, 2011). As I observed in my fieldwork, data centre and lithium activists were defending water not just as a resource they owned but instead as an agent present in their everyday life and as an enabler of their way of being. This form of resistance, I argue, is developing an *elemental ethics* that opposes the transformation of the elements into taken-for-granted AI infrastructure and encourages care and obligations towards land and communities. As I will explain, the elemental ethics mobilised by MOSACAT and the Council also raises deep questions over wellbeing and survival, questioning the prominence gained by concerns over hypothetical 'rogue' AI scenarios that could bring about human extinction.

The rest of this article is structured as follows. First, I analyse the main concerns of research on AI ethics and sustainable AI and foreground some of their limitations. After that, I turn to environmental philosophy to unpack elemental thinking. I then describe my methodological choices and procedures (interviews, participatory methods and material-semiotic analysis) and introduce the two case-studies at stake: a Google data centre project in Santiago and lithium extraction in the Atacama Desert. My analyses identified five central points underlying the resistance undertaken by MOSACAT and the Council: elemental framing, elemental injustice, elemental ignorance, elemental disruption and elemental threat. In my discussion I spell out how an elemental ethics can shift the design and development of AI. In the conclusion I summarise the main points and connect them with debates on literacy and regulation and suggest further avenues for research on this topic.

## 2. AI Ethics, Sustainability and Extinction

AI is an elusive term used to denote computational systems seeking to emulate human intelligence. While the concept has been circulating for decades, there has been a renewed interest in recent years due to the development of new techniques such as deep learning as well as the increased availability of data and computing power (McQuillan, 2022). Commonly, AI is framed in abstract and speculative terms. In practice, however, AI systems, which underly applications ranging from social media recommendation algorithms to automated weapons, rely on material resources and forms of human labour obtained from different regions of the world (Crawford & Joler, 2018). When taking this infrastructure into account, relevant social and environmental issues come to the fore such as the continuous extraction of human life through data (Couldry & Mejias, 2019), the extraction of minerals financing armed conflicts in the Democratic Republic of Congo (Taffel, 2015), the exploitative outsourcing of data work in places such as Venezuela (Posada, 2022) and e-waste dump areas ranging from Silicon Valley to Africa and Asia (Gabrys, 2011).

In recent years, the field of AI ethics has emerged as one of the main platforms for discussion around the risks posed by AI. Guidelines issued by corporate players, academic institutions and international organisations have tended to emphasise themes such as accountability, privacy and fairness. Even though these are relevant concerns, they nonetheless represent a quite limited set of impacts that, by focusing on issues that can be technically fixed, can leave unchecked relevant structural injustices underpinning the AI industry (Attard-Frost & Widder, 2023; Ricaurte, 2022; Bender et al., 2021; Powell et al., 2022; Rességuier & Rodrigues, 2020). As a consequence, questions over 'the wider contexts and the comprehensive relationship networks in which technical systems are embedded' (Hagendorff, 2020, p. 103) have remained in the background. In fact, even well-intentioned 'AI for good' initiatives have been proven to yield short-term benefits at the expense of the root causes underlying social and environmental injustices (Aula & Bowles, 2023; Madianou, 2020). The same lack of attention to both structural asymmetries and AI's value chain has also permeated the influential 'AI safety' agenda revolving around hypothetical cases of 'rogue AI', namely robots surpassing human intelligence and acting autonomously, posing existential threats (Hendrycks et al., 2023). Crucial in this regard have been ideologies circulating in the centres of AI development in the United States such as long-termism and effective altruism that have provided an excuse for an elite to impose their own priorities to the detriment of the concerns held by a broad range of other actors affected by AI (Gebru & Torres, 2023; Kaspersen & Wallach, 2022).

Studies looking at the relationship between AI and the environment, known as *sustainable AI*, also show relevant gaps. In this case, an excessive focus on the development of 'green' AI applications has tended to come at the expense of broader questions about the environmental impact of AI itself (Vaughan et al., 2023; Brevini, 2021; Van Wynsberghe, 202). In contrast, studies exploring the environmental footprint of AI infrastructure and its value chain have started to gain traction (e.g., Ligozat et al., 2022). Part of this research has sought to provide global quantitative accounts of carbon emissions, though a consensus has not been reached on whether environmental harms are growing, remaining steady or decreasing (Pasek et al., 2023). Water use constitutes an increasingly relevant issue for this literature, with a recent study estimating that global AI demand may account for 4.1 – 6.6 billion cubic meters of water use by 2027, half of the entire United Kingdom's (Li et al., 2023).

Qualitative research on the infrastructure and value chain of digital technologies has documented different ways in which AI and other recent developments can intensify socioenvironmental harm. Issues such as the commodification of nature (Velkova, 2016), the technology industry's demand for minerals benefitting armed groups (Taffel, 2015), the outsourcing of environmental damage to peripheral and semi-peripheral areas (Brodie, 2020) and conflicts emerging with Indigenous communities holding non-extractive relationships with the environment (Lehuedé, 2022a) have been explored. More recently, scholars have foregrounded the rise of activist groups opposing data centres in Europe, Latin America and the US due to pollution, water use and other externalities (Rone, 2023; Lehuedé, 2022b; Hogan, 2015).

Social justice, decolonial, feminist and Indigenous frameworks have contributed to debates on AI ethics by exploring structural and historical asymmetries underpinning the design, development and application of AI systems. Such proposals have shown how AI systems can perpetuate violence at scale by operating at the macro (e.g. automated military power), institutional (e.g. government automation) and micro (e.g. bodies and subjectivities) levels, making it necessary to think of an ethics that can speak to, rather than ignore, the majority world (Ricaurte, 2022). The lack of gender and racial diversity as well as a loose allocation of responsibilities in the field of machine learning have also constituted relevant concerns (Gray & Witt, 2021). Indigenous perspectives have been particularly emphatic in unravelling AI's entanglement with the environment, as well as in expanding narrow notions of intelligence by foregrounding the importance of environmental awareness, multispecies solidarity and the affective realm (Lorencova & Trnka, 2023; Lewis, 2020). Sub-Saharan African Ubuntu ethics has been highlighted as a means to advance a less extractive and individualistic approach to the conception and application of technologies such as machine learning (Birhane, 2021; Mhlambi, 2020).

Broadly speaking, only a narrow set of actors and concerns have been deemed as legitimate in debates on the ethics and sustainability of AI, especially when it comes to the exclusion of groups located in the majority world but nonetheless participating within AI's value chain (Amrute et al., 2022; Cave & Dihal, 2023). In this article I address this limitation by not only documenting the environmental harms engendered by AI systems on those communities but also by foregrounding how their lived experience and visions can delineate an alternative ethics for AI.

### 3. Thinking Elementally

As mentioned earlier, the centrality of water, as identified in my fieldwork, connects with *elemental thought*, namely a form of engaging with the outside world via the building blocks that bind together everything that exists, either organic or inert.

In the context of the climate crisis, environmental philosopher David Macauley (2011) has advocated elemental thought as a means of fostering an environmental re-enchantment capable of claiming back nature from scientificist, managerial and extractive frameworks. In the words of Ivan Illich: 'H2O is a social creation of modern times, a resource that is scarce and that calls for technical management. It is an observed fluid that has lost the ability to mirror the water of dreams' (Illich, 1985, p. 76). Phenomenologically speaking, elements such as water, fire and air mediate humans' experience of the environment. Whereas the notion of 'nature' suggests a somewhat colossal and abstract entity, elements are ordinary, quotidian and palpable as humans interact with them in their everyday lives. This way of thinking about nature opens the possibility of 'thinking ecology anew' (Cohen & Duckert, 2015, p. 4) since it can transform the elements into an agent mediating human-nature relations. Furthermore, elemental thought can ignite original forms of practical environmental action (Macauley, 2011, p. 345), which in this article is particularly visible in relation to activism.

Foregrounding the elements can profoundly shift the study of media and technology. Rather than mere 'resources' in the background, water, air, fire and other elements can become agents actively carrying, conveying and shaping meaning (Peters, 2015; Starosielski, 2019). Furthermore, previously taken-for-granted actors can gain relevance when taking this approach. One example concerns electrons, a subatomic particle whose material properties have enabled the miniaturisation of digital technologies and, in doing so, prompted the outsourcing of data processing and the construction of data centres (Pasek, 2023). Finally, paying attention to the mineral layer of digital systems can profoundly change the scales, temporalities and actors involved as it reveals that the development of devices such as mobile phones is deeply reliant on, and can even shape, long-term geological processes (Parikka, 2015).

This article obtains inspiration from works drawing on elemental thinking to render visible different forms of social and environmental harm and activism underpinning the value chain technology. For example, relevant injustices can take place in relation to minerals when their extraction seeks to supply the demand of the technology sector. A relevant example concerns the mining of the coltan, cassiterite, wolframite and gold ores required to build 'smart' devices and whose profits have been employed to finance armed conflicts in the Democratic Republic of Congo (Taffel, 2015). Not only harms, but also generative forms of mobilisation and solidarity can give rise to confront technology-related elemental extraction. Mél Hogan (2015) studied digital rights and environmental activists opposing the construction of a water-intensive US intelligence's data centre project in Utah. Much more than only issues of access to water were at stake in this case: 'Rather than merely cooling the servers on which our digital data rests, water holds a poetic, a politic, and a philosophy about life: who gets to live it, how it is made manifest, recorded, and archived' (2015, p. 7).

The cases of data centre and lithium activists I explore in this article connect with broader environmental justice struggles resisting water extraction in different regions of the world (Boelens et al., 2018). In the context of climate crisis, water deprivation and insecurity are affecting poorer communities the most, who at the same time are excluded from the management and governance of this element. Paradoxically, some have proposed the use of digital technologies and AI solutions to address water insecurity issues. However, such initiatives can indeed further inequality and exclusion – the case of the Colorado River in the US shows that datafication and machine learning projects can concentrate decision-making in the hands of powerful technology companies at the expense of local communities (Dryer, n.d.).

The way I approach the elements in this article is one that acknowledges their ontological plurality, namely one in which elements such as water and fire *are* different things. In fact, for different groups the elements can be approached 'as chemical categories, as cosmological forces, as material things, as social forms, as forces and energies, as sacred entities, as experimental devices, as cultural tropes, as everyday stories and as epistemic objects' (Papadopoulos et al., 2021, p. 5). Ontological openness makes it possible to value the range of cultures and philosophies that have engaged, and are still engaging, with the elements in distinct ways, such as ancient Greek philosophy, Taoist thought in China and many Indigenous cosmologies across the planet (Furuhata, 2019). In the case of developments such as AI, an ontological approach makes it possible to understand how the expansion of digital infrastructure and value chain can involve the imposition a particular *water* (e.g. as a mere resource to power technology-fuelled development) that aligns with the visions and interests of big tech companies.

In sum, a focus on the elements can foreground previously overlooked forms of socioenvironmental injustice and activism related to technological development. As I show next, sustaining and fostering an elemental ethics has been central in the activism of communities resisting the environmental impact of AI's value chain.

4. **Methods**

The empirical data in this article comprise semi-structured interviews, workshops and protest material collected between March and April 2022. When it comes to MOSACAT, I held one in-depth interview with three members of this group in their headquarters in Cerrillos (Chile) and collected protest material available on their Facebook page. I also co-organised a 'popular workshop' on 2 April 2022 with MOSACAT members titled *Data and Extractivisms* and joined MOSACAT in the World Water Day protest on 2 March of 2022 in Santiago. When it comes to the Council, I held three semi-structured interviews with a former leader, a current member of the Council's research group and a teacher who has voiced their critical views. In this article I also analyse a video available on YouTube of the Council's participation at the COP 25 conference held in Madrid in 2019 (El Correo del Sol, 2019).

Having defined water as the central concern of this study, I then conducted a thematic analysis to identify the sub-themes through which water was articulated in the data I collected.

This procedure was followed by a material-semiotic analysis of interview quotes and graphic protest material as texts carrying meaning that can both grant or obscure the agency of non-human actors such as water (Law, 2008).

My positionality in relation to the groups I collaborated with is one of insider-outsider. I grew up and did my bachelor's degree in Santiago, Chile, before pursuing an academic career in the UK. While I did not know anyone from MOSACAT before fieldwork, I have been in touch with members of the Lickan Antay community since 2019 due to my previous research on astronomy data governance in Chile. As a non-Indigenous person, some members of the Lickan Antay communities identified me as a 'Chilean'.

Even though interviewees agreed on using their real names in this research, I have decided to anonymise them in light of the changing political situation in Chile in which far-right groups opposing environmental and Indigenous rights have gained popularity and power.

The analytical section is structured in three parts. It first introduces the two case-studies and groups studied (section 5) and then unpacks five central points mobilised by MOSACAT and the Council: elemental framing, elemental injustice, elemental ignorance, elemental disruption and elemental threat (section 6). The discussion section draws connections between the elemental activism of these groups and the ethics and sustainability of AI (section 7).

## 5. Case Studies: Data Centre and Lithium Activism in Chile

As mentioned earlier, this study comprises two case-studies in Chile: a Google data centre project in Cerrillos, Santiago, and lithium extraction in the Atacama Desert. In both cases I focus on the activism undertaken by members of those communities whose views might not necessarily be shared by the entire communities.

### 5.1. *MOSACAT and the Google Data Centre Project*

MOSACAT was formed in 2019 when a group of residents in the working-class area of Cerrillos, in Santiago, heard about a Google data centre project locally. Despite their lack of technical expertise, MOSACAT members, most of whom were women, assessed the environmental report submitted by Google and found out that the data centre was planning on utilising 169 litres of water per second in an area facing drought (Arellano et al., 2020). Concerned about this situation, MOSACAT initiated a campaign to raise awareness towards the implications of the construction of the data centre. This was not an easy task as local people, especially younger generations, felt proud that a renowned company such as Google had chosen Cerrillos for this project. Furthermore, the right-wing President of Chile, Sebastián

Piñera, supported the project, and the socialist mayor of Cerrillos, Arturo Aguirre, did not initially oppose it.

Against this backdrop, MOSACAT put up posters, distributed flyers in street markets, collected signatures, carried out rallies at the construction site, organised local assemblies and held meetings with representatives of Google in order to effect change. A few months after its formation, MOSACAT managed to include a question on the data centre in a public (although non-binding) referendum, with 38% of residents approving the project and 49% rejecting it. Three years later, it laid bare that Google was committed to 'not utilising water' to cool off its servers (MOSACAT, 2022). As of December 2023, construction works had not been initiated.

*5.2.    The Council of Atacameno People and Lithium Extraction*

The second group I will consider is the Council of Atacameno People, an association made up by eighteen Lickan Antay communities inhabiting towns and villages close to the Atacama Salt Flat in the Atacama Desert. Having been annexed by the Incan Empire and Bolivia in the past, these communities have been part of Chile since 1883. Even though lithium extraction started more than fifty years ago, there has been a boom in the last decade (Jerez et al., 2021). This is because lithium is employed for the construction of rechargeable batteries powering 'intelligent' devices such as the Alexa voice assistant and the iPhone smart phone. Such devices not only support AI applications but also collect the data employed to train the algorithms that feed AI.

As of today, Chile represents the second largest source of lithium worldwide after Australia. In Chile, lithium is extracted by pumping brine pools that are left to evaporate on the surface. Unlike the rock extraction method employed in other regions, extraction from brine pools involves the use of vast amounts of water in one of the driest regions of the world. While the government presented a new National Lithium Strategy in 2023, local communities criticised its lack of participation (diarioUchile, 2023). As of today, the Chilean state permits the extraction of lithium by two companies: SQM (Chilean and Chinese capital) and Albemarle (US capital). Even though the water employed by lithium extraction is not drinkable, it is crucial for sustaining local ecosystems. In fact, this situation is already affecting species such as flamingos and algarrobo trees in the area, as well as undermining a rich microbiological diversity (Bonelli & Dorador, 2021; Tapia & Peña, 2020).

Importantly for the purpose of this article, water is a sacred element for the Lickan Antay, a tradition that has allowed them to thrive in the desert. Centuries ago, local communities had an adaptable settlement regime based on water availability. These communities still practice rituals such as '*limpia de canales*' (channel cleaning) to sustain good relations to water and land (Bolados García & Babidge, 2017). It is noteworthy that, despite the fact that the Lickan Antay communities are struggling to recover their Kunza language, *'puri'*, which translates as 'water', is one of the few words that have survived.

## 6. Mobilising Water to Mobilise Communities

In the following sub-sections, I unpack five central points of the resistance undertook by MOSACAT and the Council. I include the term 'elemental' in them since, even if this might sound repetitive, each point mobilises different articulations of this term.

### *6.1. Elemental Framing*

An elemental ethics begins by proactively situating water as *the* central concern for local communities.

While the construction of a data centre can bring about different issues, MOSACAT decided to focus its struggle on water rights after it heard about Google's project. The reason for this choice became clear to me when I asked activists whether they would be interested in forging alliances with digital rights groups already fighting against Google and other big tech companies. This is how one of them replied:

*I've heard about* 'Derechos Digitales' *[Chilean digital rights organisation], but we didn't include that aspect in our struggle … Water impresses people more quickly… The other thing [about digital rights] is like 'you are seeing ghosts where there are none' … It's easier for them to understand and feel the effects on water. You will not be able to flush the WC, have a shower, wash your clothes.*

Mobilising people around concerns such as privacy has been a long-term challenge for digital rights organisations as such harms do not easily map onto the everyday lives of ordinary people (Carmi & Nakou, 2023). Instead, bringing water into the picture can prompt an immediate response from people who do not necessarily have an in-depth understanding of the social implications of technology. This is because ordinary people interact with water in their daily lives, which transforms this element into a familiar and quotidian actor. The allusion in the above quotation to toilet-related interactions, such as flushing the toilet and having a shower, evokes the intimate relationship that people hold with this element. Furthermore, the impact of the data centre via the lens of water is something that not only can be understood but also *'felt'* – it appeals to both reason and affect, making people more prone to fully grasp what is at stake.

Similarly, the Lickan Antay communities focused their struggle on the vast amounts of water used by the lithium companies operating in the Atacama Salt Flat. As Jorge Ramírez, a member of the Council's research team, explained to me:

> *Water is very important for the people ... The lithium boom is happening because it can be used for batteries and to store energy. We see that differently here because the water required to have lithium is extracted from here. They call it 'water mining' here. Not lithium mining, you see? Vast amounts of water are extracted from here. Almost two thousand litres per second of water.*

For the communities living close to the Atacama Salt Flat, lithium mining equals water mining. This idea suggests that the AI industry is also a *water* industry considering the central role, and scarcity of water required to cool off data centres and obtain the minerals that make up AI devices. While AI companies are usually depicted as producers of code, algorithms and applications, an elemental approach incorporates the extraction of water and other basic components of life as crucial for AI companies' operations. The fact that such an extraction is outsourced does not make these operations less important to their business models.

The following quotation is also relevant when it comes to Lickan Antay communities resisting lithium extraction. This is how Ramírez explained to me the differences between the local communities and a delegation from the German government interested in increasing lithium extraction:

> *The final objective of Germany is to make this extraction sustainable. In other words, to take charge of the whole value chain. But in this value chain, the main feature is water extraction. No solution is provided for that. Solutions are given so that we are all ok. Then, from outside it* looks like *they are working on sustainability.*

The above quotation suggests that relevant power dynamics are manifested in relation to the value chain of AI. Such power dynamics speak to a hierarchisation of the different layers and processes at stake. For the representatives of the German government, sustainability is not mutually exclusive with water extraction. For the local communities, instead, the value chain will not become sustainable unless water extraction is addressed as a central concern. In the terms of the ICT industry, actors with vested interests in lithium mining can utilise vague and ambiguous understandings of 'sustainability' to deflect, rather than address, the environmental impact of technology (Vaughan et al., 2023). AI companies can promote one aspect, such as their support for 'green' technologies, to obscure their environmentally problematic footprint. In this vision, being a 'sustainable' AI company is not incompatible with resourcing water from Indigenous territories in Chile facing drought.

Centring elements such as water in environmental struggles within AI's value chain makes it possible to situate the struggle around to the everyday lives of ordinary citizens, appealing to both the intellect and emotions in order to prompt mobilisation. Whereas there are different issues emerging from living close to a data centre or with lithium extraction, choosing water as a central concern becomes a political decision and a form of resistance, as it contradicts the priorities of the AI industry and the companies resourcing it. An elemental ethics thus resists the tendency to take for granted the elements that make up AI, bringing to the fore what is usually referred to as 'infrastructure' and left in the background.

### 6.2. Elemental Injustice

Having framed the main issue at stake as one concerning water, both MOSACAT and the Council portrayed the extraction of this element as involving deep injustices.

Notions of elemental injustice are particularly salient in the protest material that MOSACAT shared on Facebook. The rich symbolism and multi-layered character of the following graphic makes it particularly interesting in analytical terms:

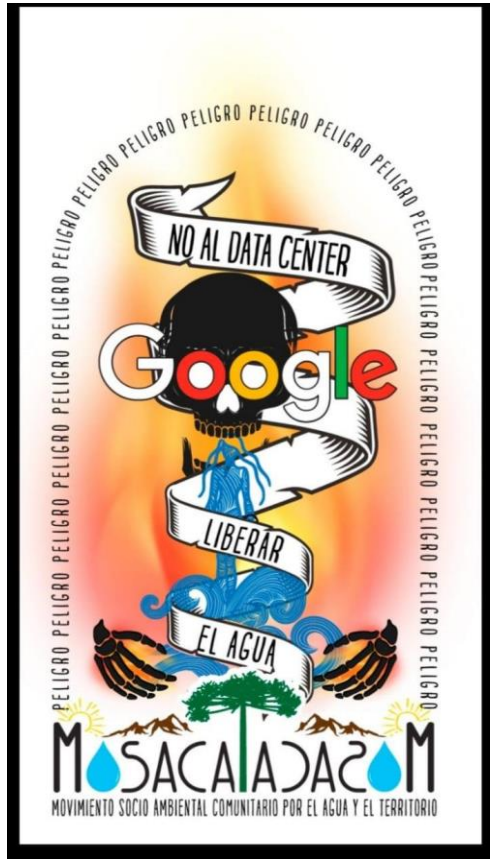

Material developed by Pamela Ramírez. Used with permission.

Source: (MOSACAT, 2020b).

At the centre of the above graphic, the two 'o's of Google's logo represent the eye sockets of a skull. Skulls are usually employed in toxic warning sings; by echoing this symbolism, MOSACAT conveys the eventual toxicity of Google's data centre project and the idea that it can directly affect the wellbeing of the community. Below the skull, what MOSACAT activists call the 'water woman' emerges from waves. The hands of the skull wrapping the water woman links to a signifier that was frequently used by activists – *appropriation*. In fact, one of the slogans they employed both in Cerrillos and in the World Water Day protest I attended was *'no es sequía, es saqueo'* ('this isn't a drought, this is a robbery'). The central stripe of the graphic contains something akin to a papyrus that reads 'No to the data centre. Free up the water'. This slogan suggests that MOSACAT rejects appropriation not only because the water *belongs* to the community (which would be a mere ownership issue) but also because this appropriation does not let water be free. In this way, activists are situating the dispute as concerning not only distribution but also as an ontological issue, i.e. as a controversy revolving around what the elements are, need and can do.

This anthropomorphisation of water and attribution of rights (freedom) represents an original form of providing agency to non-human actors enabled by elemental thought (Papadopoulos et al., 2021). At the same time, it echoes the Latin American notion of 'rights of nature' (Escobar, 2018, p. 147) through which local and Indigenous communities have challenged anthropocentric state formations, making the law compatible with worlds in which elements such as water not only need to be protected but also constitute political agents in themselves. Building upon MOSACAT's vision, an elemental ethics would not only bring to the fore but also provide agency and voice to water and other elements in the design and governance of AI.

Like MOSACAT, the Council has also approached water extraction as one concerning an elemental injustice. As they explained to me, the reason why lithium companies privilege Chile lies in the simple fact that extracting lithium in that country is cheaper and less regulated. However, the injustices regarding lithium also have an additional layer of complexity as, despite the harms engendered by its extraction in the Atacama Desert, lithium is currently considered as a key resource for the transition towards post-carbon societies. This is evident in the following words by Ramírez:

*We cannot be traded for the energetic transition … We cannot be a sacrifice zone. Unlike a sacrifice zone, which is very easy to see and has an immediate impact, this [the Atacama Salt Flat] can be a sacrifice zone in the long term. Because there is a slow death happening here. Slow. Slow. Slow. But there has been an accumulation already. Besides, that accumulation will be extensive. The underground water, the aquifers, can take one hundred years or more to recover. Imagine that.*

The concept of 'sacrifice zone' portrays an injustice whereby peripheral areas assume most of the costs of sustaining the core's consumerist way of life (De Souza, 2021). But in addition to this distributive form of injustice (unequal access to water), lithium extraction also encompasses an injustice whereby a group is able to impose their ecological views and design for transition towards low-carbon futures. For some, lithium constitutes a 'transition mineral' as it is considered key for a technologically-supported transition towards green and sustainable economies. This discourse has been employed as a justification for lithium extraction in Chile, being adopted not only by external actors but also by local authorities as well (Voskoboynik & Andreucci, 2022, p. 799).

The above quotation also points to the different temporalities that, in the vision of Ramírez, should be taken into consideration. When referring to the hundreds of years that can take to recover the underground water, Ramírez argues that short-term human-centred temporality needs to be complemented with a long-term geological one. In this way, an elemental ethics embraces the 'conjoined histories' (Chakrabarty, 2021, p. 49) tying up together humans and the elements in the designing and governing AI.

Thus, both MOSACAT and the Council criticise the fact that local communities and environments can pay the price of technological progress and seemingly 'green' and 'sustainable' ecological visions. But beyond this distributive point, an elemental ethics also opposes extractive logics that undermines the way these communities relate to water and the environmental. Furthermore, the elemental ethics of MOSCAT and the Council reveals that injustices taking place within AI's value chain can not only undermine communities and humans but also the agency, needs and voice of the elements themselves.

*6.3. Elemental Ignorance*

A third point present in both MOSACAT's and CAP's resistance concerns the lack of information regarding the availability of underground water.

MOSACAT formed after Cerrillos residents found out that the Google data centre project would utilise 169 litres of water per second, which equals to one day of consumption of the more than 80,000 Cerrillos residents (Arellano et al., 2020, p. 201). This information ended up becoming crucial in MOSACAT's public interventions and protest material, as exemplified in the following poster put up at a bus stop:

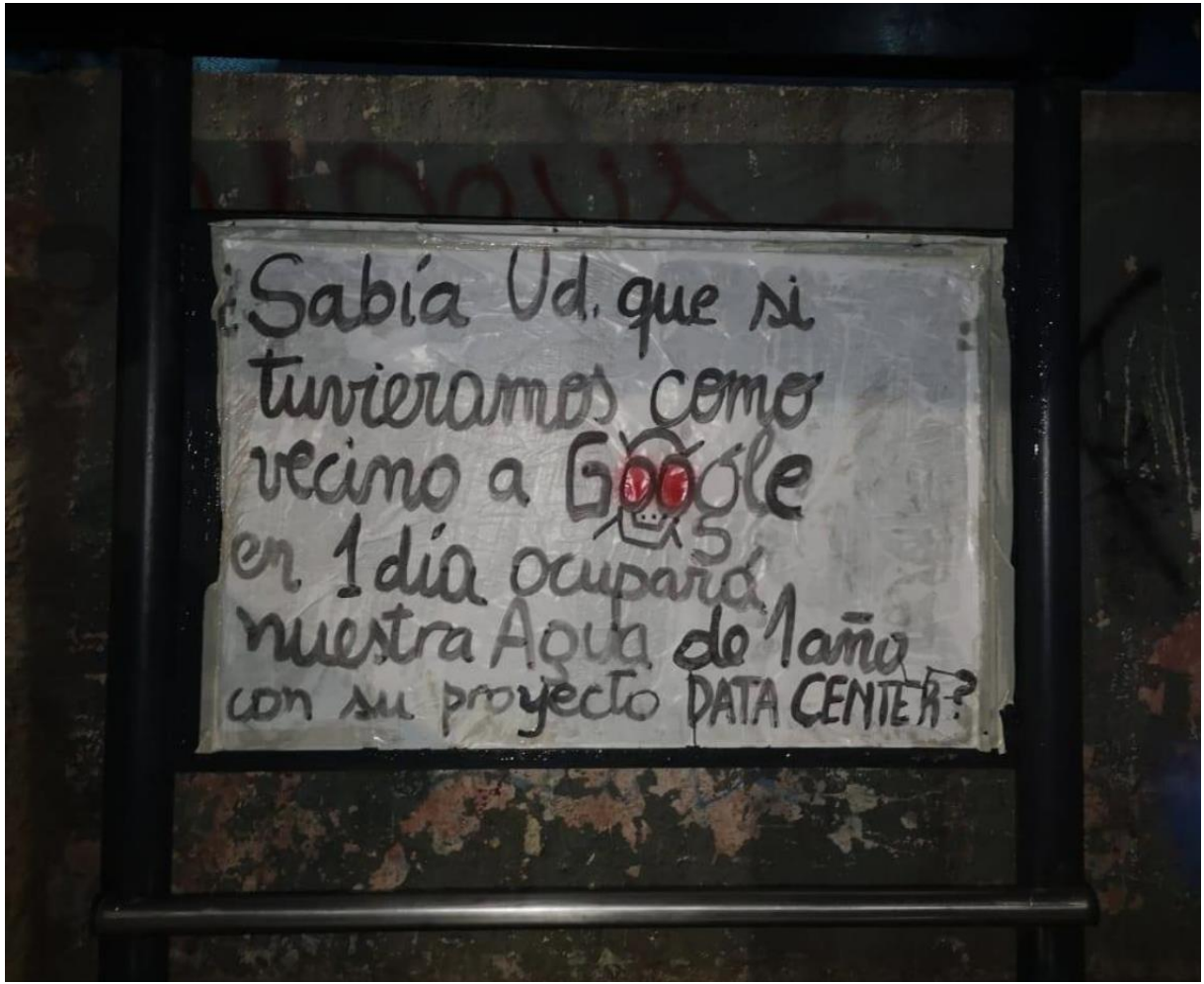

*'Did you know that if we would have Google as our neighbour,*
*in 1 day it will use our yearly water with its data centre project?'.*
Material developed by Pamela Ramírez. Used with permission.
Source: (MOSACAT, 2020a)

But despite having been able to uncover this statistic, there is a crucial piece of information that has been omitted: the overall availability of the underground water supplying Cerrillos. This lack of information, which could have vital consequences for the everyday life of residents, was strategically mobilised by MOSACAT:

*I would shock them [the residents]. I would tell them: 'Imagine that they will come to deliver water to you on a water truck'. And that is shocking for anyone ... So I would tell them: 'They will cut the water so that a company can keep operating. They will cut our water, imagine, from 3 to 4 in the afternoon'. We do not have certainty.*

After the construction of Google's data centre, Cerrillos could become just one among several other municipalities in Chile served by water trucks in the context of unprecedented drought. To address this risk, one of the main demands by MOSACAT has been a hydrogeological study that could determine the availability of water in the area. Unfortunately, such a request has not been addressed by either the municipality of Cerrillos, the municipality-owned water provider SMAPA or Google itself.

The same concern is shared by the Lickan Antay communities in the Atacama Desert. These local communities believe that mineral extraction in the Atacama Salt Flat is using 4,100 litres of water per second. However, just like in Cerrillos, there is no overall information regarding the availability of the underground water supplying the area.

A former leader of the Council, Francisco Valdivieso, referred to this point in the following terms:

*The ideal situation would had been one in which they would have done a study to assess the capacity of the salt flat to keep being exploited in the way it is now. This study does not exist. We have a sick person, but a sick person who has not gotten any exam to check their status. It's unknown whether this person's status is recent, getting into an illness, in the middle of it, or if this is a terminally ill patient. We do not know that.*

There is, however, evidence of the damage brought about by lithium extraction in the Atacama Desert. The decrease in the population of flamingo colonies and algarrobo trees as well as the loss of microbial diversity are examples of this. Lickan Antay communities do not have enough information to figure out whether this is the beginning or the end of something, and to what extent the Atacama Desert could still support lithium extraction. In fact, in 2019 a Chilean environmental court conceded that there was 'scientific uncertainty' (Primer Tribunal Ambiental, 2019, p. 266) concerning water availability in the area. A hydrogeological study would be a crucial first step to shed light on these questions.

Despite possessing basic information on water use, both MOSACAT and the Council are struggling to calculate the amount of water available in their areas. While the AI industry is happy to invest considerable amounts of resources in the development of so-called 'green' AI applications, communities and environments affected by the AI value chain are being kept in the dark when it comes to water availability. A form of strategic ignorance (McGoey, 2019) is taking place, as it might be the case that, if more situated and grounded details on its community and environmental impact were made available, the operations and profits of the AI industry would be affected on a global scale. Furthermore, elemental ignorance also raises an important to question as to whether already existing evidence should suffice in order to take action. A 'threshold theory of pollution' (Liboiron, 2021) seems to be holding sway in data centre construction and lithium extraction that tolerates environmental damage (such as the death of flamingos and algarrobo trees) as long as such damage is not deemed irreversible by quantitative scientific standards. The elemental ethics held by activists, instead, is not waiting for an environmental tipping point to be reached in order to care and act.

*6.4. Elemental Disruption*

While both Cerrillos and the Lickan Antay communities are facing threats, there is a type of disruption affecting the latter group that is worth discussing due to its elemental character.

Water represents a sacred element for the Lickan Antay people. Valdivieso's answer to my question on the difference between the Lickan Antay's and the lithium companies' approach to water is worth reproducing in toto:

*Water is essential for the Lickan Antay people. It is the resource that allows you to live. There is a close relationship between the Lickan Antay people, the water, land, the sacred mountains. In that world vision, water is recognised as the veins of Mother Earth. And water for our people is essential. It is sacred. We are in one of the driest deserts of the world, but despite that, water, in the way the elder would considerably take care of it, made it possible to have the villages that we have today ... The '*limpia de canales*' [channels cleaning], a costume of the villages, includes a ritual to the water ... That hopefully there will be more water. That hopefully there will be more rain. That we will be able to grow our vegetables, our corn, our fruit. That we will be able to keep this green. That we will be able to live.*

The above quotation portrays the particular relationship of the Lickan Antay people with water. Water is elemental in that it is 'essential' (a word that Valdivieso employs twice) for sustaining life and the Lickan Antay way of living. This relationship is sustained and kept alive through 'channels cleaning', a ritual that mobilises entire towns during 2 or 3 days so that the water can flow more easily from the mountains to the communities. This ritual ensures a good relationship with the community's ancestors, the Pachamama (land) and the mountains that protect the communities (Bolados García & Babidge, 2017).

The coexistence of ancestral and modern visions of water is prompting profound questions for the Lickan Antay communities. In particular, there are internal divergences in opinion regarding the approach that the local communities should take when it comes to the monitoring and study of water by the Council in order to hold lithium companies accountable. This is how Ramírez explained to me this situation:

*I did some interviews to the Elder ... They would tell me: 'Ok, that's fine, but what's the point of monitoring, of doing those studies, if, in the end, you are still intervening. For me, the ideal would be that these ponds would be shut off and that no one could go in'. I explain to them [the Chilean state and lithium companies] that the elder think in that way, you know? ... Obviously, I would rather study it, you know?*

Integrating both ancestral and scientific knowledge has been crucial for the Council's research team given the two approaches to water at play. For the older community members, monitoring water, which involves the use of scientific equipment, does not constitute a proper way of controlling lithium companies as it reproduces what they deem as a problematic managerial approach. Ramírez grew up in one of the communities but also has a degree in engineering, so part of his work has been to reconcile ancestral and scientific approaches to water.

The extraction of water encompassed by lithium mining represents a disruption of the elemental relationship that the Lickan Antay people hold with water – in other words, an *elemental* disruption. This disruption has an ontological character as it pertains to profound differences regarding what water is and wants – a sacred element? An agent that deserves protection? A resource to be managed?

*6.5. Elemental Threat*

A final point raised by both MOSACAT and the Council concerns the extent to which data centres and lithium extraction encompasses a threat to the wellbeing and even survival of communities.

In the case of MOSACAT, there are concerns that the Google data centre could end up leaving the community without access to pipeline water. This situation generates frustration for MOSACAT. As one of them expressed to me:

*It is sad to see that you are worth nothing. That your humanity is not considered. And that they make business out of your lacks. How could that not upset us. It upsets us and puts us in a position of class struggle between those who have the power and are squeezing us, and those of us who are right but that do not have that economic power.*

The above quotation criticises how profit-seeking can prevail over the wellbeing of the Cerrillos residents. This situation is making the local activists approach the issue as class struggle between those at the top and those at the bottom of the value chain of digital technologies. In a landscape where such activities situate MOSACAT within this structure, there is no single 'humanity' but a hierarchisation of different 'humanities'.

On the other hand, the Lickan Antay communities have been explicit in the connection between lithium extraction, digital technologies and extinction scenarios. This was especially the case in the 2019 UN Climate Change Conference COP 25 in Madrid (El Correo del Sol, 2019), when Valdivieso gave the following speech:

*We are leaving people without water to take the water to the lithium operations and to be able to enjoy these electronic devices that we all enjoy nowadays ... We will go to great lengths to show that our people today are close to extinction due to mining.*

The above quotation denounces the existential threat faced by the Lickan Antay people, one tightly connected to the mass-scale production of devices such as mobile phones. However, the communities near the Atacama Salt Flat are willing to resist and to make as much noise as possible in order to preserve their wellbeing. It is noteworthy that Valdivieso referred to 'extinction' to highlight the harms taking place within the AI value chain before this word was mobilised by AI experts and pundits, most of them based in the Global North, to depict hypothetical scenarios brought about by AI (Hendrycks et al., 2023).

*Cultural* extinction, namely the extinction of local visions and practices, is also a concern for Lickan Antay communities. For school teacher Oriana Mora, the Lickan Antay way of living is tightly linked to the wellbeing of territory and water. In her words:

*I am not sure if I could be who I am elsewhere. I mean, I have a relationship with the territory. In this case, if I have a relationship with the mountains, with the channels to which I carry out ceremonies. I am not sure if I could go somewhere else to do that. So I surely depend on this territory ... So it is surely unthinkable for us to not live here.*

According to the above quote, the Lickan Antay are who they are because of the specific configuration and relationships enabled by the territory. Such a *relational* ontology regards individuals, communities and the elements as mutually constitutive and dependant on each other, contrasting with the dualist vision that emerged in modernity that sees nature as the mere background to culture and society (Escobar, 2018, p. 95). As long as an extractive approach is privileged, extraction taking place within the value chain of technology can intensify the separation experienced by local communities from the land in which they live.

As I have shown, both MOSACAT and the Council have articulated different threats posed by water extraction. In this case, the elemental of elemental ethics not only points to the elements that make up the world but also to what is fundamental for sustaining life, namely *vital*. Both these groups posed this situation as not only a life- and culture-threatening issue but also as one that is hierarchising the wellbeing and survival of some 'humanities' over others.

Paying attention to the elements reveals how discussions over AI extinction appeal to *hypothetical scenarios* brought about by 'rogue AI' are distracting from the voices of communities facing concrete risks. As the extinction of some groups becomes sidelined, a form of *necropolitics* creates differentiated categories of humanities and dictates whose forms of life matter and deserve protection (Mbembe, 2019).

### 7. Discussion: Outlining An Elemental Ethics for Artificial Intelligence

Conceptually speaking, this article shows how elemental thinking as discussed in environmental and science and technology studies can speak to the needs and visions of communities resisting the environmental impact of AI. More specifically, an *ontological* approach to the elements enables a fine-grained and grounded approach to the material and symbolic harms encompassed by AI. At stake, I have shown, is not only environmental degradation but also the disruption of the relations through which communities experience and take care of the environment and its elements. In the terms of MOSACAT activists, the elements' freedom to act as agents is at risk. Contributing to literature on elemental media and technology, I have demonstrated that elemental thinking can deepen injustices taking place at the level of infrastructure and value chain, as well as mobilise ordinary citizens against those injustices.

An elemental ethics also shows how ontological disputes are gaining visibility in the design and development of technology. This is not a coincidence since divergences over what exists and what does not, over what counts as Life and what does not, have become paramount in the context of late liberalism and the climate crisis (Povinelli, 2016). The cases of MOSACAT and the Council reveal that ontological disputes are also gaining prominence in the field of AI via conflicts taking place in the value chain. Inspired by centuries of territorial struggles, the ontological plurality sustained by an elemental ethics is known in Latin American decolonial thought as the *pluriverse*, or 'a world in which many worlds fit' (Escobar, 2018, p. 16). In the context of the climate crisis, disputes of this kind are deemed to intensify as technological developments are used as an excuse to impose techno-centric ecological visions on local communities.

Beyond theory, the cases of MOSACAT and the Council can inspire an ethics of resistance that would centre elemental relations in the design and governance of AI. Indeed, the first aspect of an elemental ethics comprises an acknowledgement of the crucial role that the elements – water in the case of this study, but also other ones such as air or fire – have for the development of intelligent systems and for the communities partaking within AI's value chain. For MOSACAT and the Council, the elements are not mere resources but also agents present in their everyday life as well as enablers of their ways of life. Bringing the elements to the fore resists the modern tendency to take for granted, and to set in the background, the components that make up technology– in other words, this ethics challenges the transformation of the elements into mere *infrastructure*. Given the water-intensive character of AI, an elemental ethics calls for acknowledging the extraction of resources as a crucial aspect of what technology companies do and the ways in which they obtain profits, challenging representations of developments such as AI as abstract and immaterial ones.

A focus on elemental relations can also expand approaches to AI sustainability. While quantitative measurements over the use of water and other elements can help provide a general picture, they do not specify how these harms manifest in specific settings. Situated, empirical and qualitative studies attending to the needs and visions of the communities and environments participating within AI value chain are required. Based on MOSACAT and the Council's activism, an elemental approach can help understand how resource extraction can affect the unique relationship and dependencies that communities hold with their environments. Furthermore, an elemental approach shows that universal and top-down approaches to 'greening' AI can fail to acknowledge the very fact that elements such as water *are*, and therefore *need* and *want*, different things in different contexts. In addition to this, an elemental ethics to AI would mobilise both reason and affect, expand the temporalities at stake (not only human but also geological ones) and critique modern science's imperative to master nature.

A crucial point in the discussion over sustainability concerns issues of accountability and the distribution of responsibilities. In this regard, an elemental ethics proposes a shift towards *situated* obligations according to the role that different actors hold within the vast network of human and more-than-human interdependencies that make AI possible. As I explained earlier, the prevailing form of moral reasoning in the field of AI draws on utilitarian ethics focused on the maximisation of benefits. Taken to an extreme, this approach can justify environmental harm in a specific area in the name of a quantitatively-measured and centrally-identified greater good. In contrast, an elemental ethics highlights the importance of understanding the position that companies and other actors hold within the whole value chain of AI and how this value chain can affect elemental relations in specific contexts. 'Located accountability' approaches (Widder & Nafus, 2023), rather than universal imperatives and guidelines, are demanded by an elemental ethics. Because of this, much more aligned to an elemental ethics than utilitarianism is *care* ethics – an approach that focuses on the responsibilities of each actor within the network of interdependencies and invisible labour that sustain the world (Puig de la Bellacasa, 2017).

Finally, an elemental perspective shows how powerful actors have been able to prevail in discussions on the existential threats posed by AI. In both cases analysed (the Google data centre and lithium extraction), local communities have expressed relevant concerns over the impact of the value chain of AI for their wellbeing and the survival of their way of life. Unfortunately, such concerns have been excluded from the influential 'AI extinction' agenda. So far, this agenda has been dominated by voices concerned about the potential emergence of a supra-human computational intelligence, called artificial *general* intelligence, that could endanger the continuity of the human species (e.g. Bostrom, 2014). Bioterrorism, automated warfare, rogue robots and other hypothetical scenarios have been syndicated as catastrophic risks engendered by AI (Hendrycks et al., 2023). These concerns have rapidly gained prominence in academia and global governance arena, partly thanks to the lobby of organisations with ties to the AI industry (Bordelon, 2023; Tiku, 2023). From a historical perspective, the undermining of Indigenous visions of social and environmental apocalypses underpinning the field of AI is not a coincidence but a recurrent phenomenon that still informs mainstream environmental thinking (Vilaça, 2023).

Just to make clear, an elemental ethics does not propose a yes/no approach to AI development. Instead, elemental ethics insists that no technological system can be deemed as 'intelligent', 'ethical' or 'sustainable' as long as its resourcing disrupts the elemental relations sustaining the planet. Not only this, truly intelligent, ethical and smart systems should be capable of identifying what is elemental for the communities and the environments participating within AI value chain and seek ways of contributing to the flourishing of multiple ways of relating to the elements.

## 8. Conclusion

In this article I showed that no 'ethical' and 'sustainable' AI would be possible as long the communities participating within AI value chain, and their ways of relating to the elements, are excluded from the design and development of so-called 'intelligent' systems. Drawing on interviews and protest material, I analysed how MOSACAT in Cerrillos and the Council of Atacameno Peoples in the Atacama Desert resisting a Google data centre project and lithium extraction respectively are mobilising water in order to mobilise their communities against extraction. I showed that their resistance is sustained in five central points: (1) the framing of the issue at stake as an elemental one, (2) an elemental injustice concerning distribution and imposition of ecological views, (3) an elemental ignorance on vital information such as water availability, (4) a disruption to their elemental relations (in the case of Lickan Antay communities) and (5) an elemental threat to their wellbeing and survival. After that, I then outlined an elemental ethics for AI, echoing approaches based on care and obligations that acknowledges situated elemental relations. As I showed, an elemental ethics deeply challenges the top-down, utilitarian and quantitative moral reasoning prevailing in debates on AI ethics and sustainable AI.

The elemental ethics I developed in this study can not only inform the way we *think* but also the way we *do* AI in terms of design, literacy and regulation. From a *design* perspective, it proposes a 'looking back' approach where those envisioning and deploying AI applications should consider not only the effects such applications can have on their users but also on the whole set of communities and environments making those applications possible. An elemental ethics also sheds lights on how the elements have been a missing piece in *literacy* proposals seeking to advance a critical engagement with AI systems. Not only algorithms and code but also minerals such as lithium and elements such as water should figure as relevant subjects in those programmes. Hopefully, the approach taken would not reduce the elements to resources but also account for the ways in which they can enable different ways of life. In terms of *regulation*, this study lays bare the importance of not only addressing the environmental impact of AI systems but also the involvement of communities who have been excluded from such discussions. Relevant links are being forged at the moment between digital rights and environmental activists; besides this necessary move, communities themselves should be included so as to let them safeguard their elemental relations.

As I have shown, an elemental ethics provides an ethics that, inspired in the resistance of grassroots and Indigenous groups, questions dominant understandings of the relationship between AI and the environment. By privileging a bottom-up approach, it contrasts the extractive approach privileged by AI industry with the multiple ways in which elements such as water come to matter for local communities. While in this article I have privileged an ontological approach, political economy perspectives could also shed light on the essential role of elements such as water in what AI companies do and how they obtain their profits.

## 9. References

Amrute, S., Singh, R., & Lara Guzmán, R. (2022). *A Primer on AI In/From the Majority World: An empirical Site and Standpoint*. Data & Society Research Institute. http://dx.doi.org/10.2139/ssrn.4199467

Arellano, A., Cifuentes, L., & Ríos, C. (2020, May 25). Las zonas oscuras de la evaluación ambiental que autorizó 'a ciegas' el megaproyecto de Google en Cerrillos [The Dark Zones of Environmental Assessment that 'Blindly' Authorised Google's Megaproject in Cerrilos]. *Ciper*. https://www.ciperchile.cl/2020/05/25/las-zonas-oscuras-de-la-evaluacion-ambiental-que-autorizo-a-ciegas-el-megaproyecto-de-google-en-cerrillos/

Attard-Frost, B., & Widder, D. G. (2023). The Ethics of AI Value Chains: An Approach for Integrating and Expanding AI Ethics Research, Practice, and Governance. *arXiv Preprint*. https://doi.org/arXiv:2307.16787

Aula, V., & Bowles, J. (2023). Stepping Back from Data and AI for Good – Current Trends and Ways Forward. *Big Data & Society*, *10*(1). https://doi.org/10.1177/20539517231173901

Bender, E. M., Gebru, T., McMillan-Major, A., & Shmitchell, S. (2021). On the Dangers of Stochastic Parrots: Can Language Models Be Too Big? 🦜. *Proceedings of the 2021 ACM Conference on Fairness, Accountability, and Transparency*, 610–623. https://doi.org/10.1145/3442188.3445922

Birhane, A. (2021). Algorithmic Injustice: A Relational Ethics Approach. *Patterns*, *2*. https://doi.org/10.1016/j.patter.2021.100205

Boelens, R., Vos, J., & Perreault, T. (2018). Introduction: The Multiple Challenges and Layers of Water Justice Struggles. In R. Boelens, T. Perreault, & J. Vos (Eds.), *Water Justice* (1st ed., pp. 1–32). Cambridge University Press.

Bolados García, P., & Babidge, S. (2017). Ritualidad y extractivismo. La limpia de canales y las disputas por el agua en el Salar de Atacama-Norte de Chile [Rituality and Extractivism. Channels Cleaning and Disputes over Water in the Atacama Salt Flat-North of Chile]. *Estudios Atacameños*, *54*, 201–216.

Bonelli, C., & Dorador, C. (2021). Endangered Salares: Micro-Disasters in Northern Chile. *Tapuya: Latin American Science, Technology and Society*, *4*. https://doi.org/10.1080/25729861.2021.1968634

Bordelon, B. (2023, October 13). *How a Billionaire-Backed Network of AI Dvisers Took Over Washington.* Politico. https://www.politico.com/news/2023/10/13/open-philanthropy-funding-ai-policy-00121362

Bostrom, N. (2014). *Superintelligence: Paths, Dangers, Strategies*. Oxford University Press.

Brevini, B. (2021). *Is AI Good for the Environment?* Wiley.

Brodie, P. (2020). Climate Extraction and Supply Chains of Data. *Media, Culture and Society*, *42*(7–8), 1095–1114. https://doi.org/10.1177/0163443720904601

Carmi, E., & Nakou, P. (2023). *What Mobilise People to Go Against Big Tech?* Department of Sociology and Criminology, School of Policy and GLobal Affairs, City, University of London, UK. https://openaccess.city.ac.uk/id/eprint/31373/

Cave, S., & Dihal, K. (2023). *Imagining AI: How the World Sees Intelligent Machines*. Oxford University Press.

Chakrabarty, D. (2021). *The Climate of History in a Planetary Age*. University of Chicago Press.

Cohen, J. J., & Duckert, L. (2015). Eleven Principles of the Elements. In J. J. Cohen & L. Duckert (Eds.), *Elemental Ecocriticism: Thinking with Earth, Air, Water, and Fire* (pp. 1–26). University of Minnesota Press.

Couldry, N., & Mejias, U. A. (2019). *The Costs of Connection: How Data is Colonizing Human Life and Appropriating It for Capitalism*. Stanford University Press.

Cowls, J., Tsamados, A., Taddeo, M., & Floridi, L. (2023). The AI Gambit: Leveraging Artificial Intelligence to Combat Climate Change—Opportunities, Challenges and Recommendations. *AI & Society, 38(1), 283-307.* https://doi.org/10.1007/s00146-021-01294-x

Crawford, K., & Joler, V. (2018). *Anatomy of an AI System*. https://anatomyof.ai/

De Souza, M. L. (2021). 'Sacrifice Zone': The Environment–Territory–Place of Disposable Lives. *Community Development Journal*, *56*(2), 220–243. https://doi.org/10.1093/cdj/bsaa042

diarioUchile. (2023, May 5). *Consejo de Pueblos Atacameños rechazó Estrategia Nacional del Litio [Council of Atacameno People Rejected National Lithium Strategy]*. https://radio.uchile.cl/2023/05/05/consejo-de-pueblos-atacamenos-rechazo-estrategia-nacional-del-litio/

El Correo del Sol. (2019, December 3). *Los atacameños denuncian el expolio del agua para extraer litio con que se fabrican baterías [The Atacameno Denounce Water Plunder for Lithium Extraction Used For Developing Batteries]*. https://www.youtube.com/watch?v=GxNDYz_Zt0E

Escobar, A. (2018). *Designs for the Pluriverse: Radical Interdependence, Autonomy, and the Making of Worlds*. Duke University Press.

Furuhata, Y. (2019). Of Dragons and Geoengineering: Rethinking Elemental Media. *Media+Environment*, *1*(1). https://doi.org/10.1525/001c.10797

Gabrys, J. (2011). *Digital Rubbish: A Natural History of Electronics*. University of Michigan Press.

Gebru, T., & Torres, É. P. (2023). *SaTML 2023—Eugenics and the Promise of Utopia through AGI*. https://www.youtube.com/watch?v=P7XT4TWLzJw

Gray, J. E., & Witt, A. (2021). A Feminist Data Ethics of Care Framework for Machine Learning: The What, Why, Who and How. *First Monday*, *26*(12). https://doi.org/10.5210/fm.v26i12.11833

Hagendorff, T. (2020). The Ethics of AI Ethics: An Evaluation of Guidelines. *Minds and Machines*, *30*(1), 99–120. https://doi.org/10.1007/s11023-020-09517-8

Hendrycks, D., Mazeika, M., & Woodside, T. (2023). *An Overview of Catastrophic AI Risks.* arXiv. https://doi.org/10.48550/arXiv.2306.12001

Hogan, M. (2015). Data Flows and Water Woes: The Utah Data Center. *Big Data & Society*, *2*(2). https://doi.org/10.1177/2053951715592429

Illich, I. (1985). *H2o and the Waters of Forgetfulness: Reflections on the Historicity of 'Stuff'*. Heyday Books.

Jerez, B., Garcés, I., & Torres, R. (2021). Lithium Extractivism and Water Injustices in the Salar de Atacama: The Colonial Shadow of Electromobility. *Political Geography*, *87*. https://doi.org/10.1016/j.polgeo.2021.102382

Kaspersen, A., & Wallach, W. (2022, February 29). *Long-Termism: An Ethical Trojan Horse*. https://www.carnegiecouncil.org/media/article/long-termism-ethical-trojan-horse

Law, J. (2008). Actor Network Theory and Material Semiotics. In Turner (Ed.), *The New Blackwell Companion to Social Theory* (pp. 141–158). Blackwell.

Lehuedé, S. (2022a). Territories of Data: Ontological Divergences in the Growth of Data Infrastructure. *Tapuya: Latin American Science, Technology and Society*, *5*. https://doi.org/10.1080/25729861.2022.2035936

Lehuedé, S. (2022b). Big Tech's New Headache: Data Centre Activism Flourishes Across the World. *Media@LSE*. https://blogs.lse.ac.uk/medialse/2022/11/02/big-techs-new-headache-data-centre-activism-flourishes-across-the-world/

Lewis, J. E. (Ed.). (2020). Indigenous Protocol and Artificial Intelligence Position Paper. *The Initiative for Indigenous Futures and the Canadian Institute for Advanced Research (CIFAR)*. https://doi.org/10.11573/spectrum.library.concordia.ca.00986506

Li, P., Yang, J., Islam, M. A., & Ren, S. (2023). *Making AI Less 'Thirsty': Uncovering and Addressing the Secret Water Footprint of AI Models* (arXiv:2304.03271). arXiv. https://doi.org/10.48550/arXiv.2304.03271

Liboiron, M. (2021). *Pollution Is Colonialism*. Duke University Press.

Ligozat, A.-L., Lefevre, J., Bugeau, A., & Combaz, J. (2022). Unraveling the Hidden Environmental Impacts of AI Solutions for Environment Life Cycle Assessment of AI Solutions. *Sustainability*, *14*(9). https://doi.org/10.3390/su14095172

Lorencova, R., & Trnka, R. (2023). Variability in Cultural Understandings of Consciousness: A Call for Dialogue with Native Psychologies. *Journal of Consciousness Studies*, *30*(5), 232–254. https://doi.org/10.53765/20512201.30.5.232

Macauley, D. (2011). *Elemental Philosophy*. SUNY Press.

Madianou, M. (2020). Nonhuman Humanitarianism: When 'AI for Good' Can Be Harmful. *Information, Communication & Society*, *24*(6), 850–868.

Mbembe, A. (2019). *Necropolitics*. Duke University Press.

McGoey, L. (2019). *The Unknowers: How Strategic Ignorance Rules the World*. Bloomsbury.

McQuillan, D. (2022). *Resisting AI: An Anti-Fascist Approach to Artificial Intelligence*. Bristol University Press.

Mhlambi, S. (2020). From Rationality to Relationality: Ubuntu as an Ethical & Human Rights Framework for Artificial Intelligence Governance. *Carr Center Discussion Paper*. https://carrcenter.hks.harvard.edu/publications/rationality-relationality-ubuntu-ethical-and-human-rights-framework-artificial

MOSACAT. (2020a, August 20). *'Volveremos, venceremos y viviremos' ['We will return, triumph and live']*. https://www.facebook.com/mosacatchile/photos/pb.100077589518337.-2207520000/338479144203674/?type=3

MOSACAT. (2020b, September 30). *'Creé este diseño' [I created this design]*. https://www.facebook.com/mosacatchile/photos/pb.100077589518337.-2207520000/373033610748227/?type=3

MOSACAT. (2022, June 21). *'Vecinos y vecinas' [Neighbours]*. https://www.facebook.com/mosacatchile/posts/pfbid02KSwiVDA9QRofxhtWM4VA4uAjazndeBpZePzchnTzW7xpnVCL5LqVRyihQAGWXnZcl

Papadopoulos, D., Puig de la Bellacasa, M., & Myers, N. (2021). Elements: From Cosmology to Episteme and Back. In D. Papadopoulos, M. Puig de la Bellacasa, & N. Myers (Eds.), *Reactivating Elements: Chemistry, Ecology, Practice* (pp. 1–17). Duke University Press.

Parikka, J. (2015). *A Geology of Media*. University of Minnesota Press.

Pasek, A., Vaughan, H., & Starosielski, N. (2023). The World Wide Web of Carbon: Toward a Relational Footprinting of Information and Communications Technology's Climate Impacts. *Big Data & Society*, *10*(1). https://doi.org/10.1177/20539517231158994

Peters, J. D. (2015). *The Marvelous Clouds: Toward a Philosophy of Elemental Media*. University of Chicago Press.

Posada, J. (2022). Embedded Reproduction in Platform Data Work. *Information, Communication & Society*, *25*(6), 816–834. https://doi.org/10.1080/1369118X.2022.2049849

Povinelli, Elizabeth A. (2016). *Geontologies: A Requiem to Late Liberalism*. Duke University Press.

Powell, A. B., Ustek-Spilda, F., Lehuedé, S., & Shklovski, I. (2022). Addressing Ethical Gaps in 'Technology for Good': Foregrounding Care and Capabilities. *Big Data & Society*, *9*(2). https://doi.org/10.1177/20539517221113774

Primer Tribunal Ambiental. (2019). *Comunidad Indígena Atacameña de Peine con Superintendencia del Medio Ambiente [Atacameno Indigenous Community from Peine with Environmental Superintendency]*. https://www.portaljudicial1ta.cl/sgc-web/ver-causa.html?rol=R-17-2019&doc=2539

Puig de la Bellacasa, M. (2017). *Matters of Care: Speculative Ethics in More Than Human Worlds*. University of Minnesota Press.

Rességuier, A., & Rodrigues, R. (2020). AI Ethics Should not Remain Toothless! A Call to Bring Back the Teeth of Ethics. *Big Data and Society*, *7*(2). https://doi.org/10.1177/2053951720942541

Ricaurte, P. (2022). Ethics for the Majority World: AI and the Question of Violence at Scale. *Media, Culture and Society*, *44*(4), 726–745. https://doi.org/10.1177/01634437221099

Rone, J. (2023). The Shape of the Cloud: Contesting Data Centre Construction in North Holland. *New Media & Society*. https://doi.org/10.1177/14614448221145928
31

Starosielski, N. (2019). The Elements of Media Studies. *Media+Environment*, *1*(1). https://doi.org/10.1525/001c.10780

Taffel, S. (2015). Towards an Ethical Electronics? Ecologies of Congolese Conflict Minerals. *Westminster Papers in Culture and Communication*, *10*(1), 18–33. https://doi.org/10.16997/wpcc.210

Tapia, D., & Peña, P. (2020). White Gold, Digital Destruction: Research and Awareness on the human Rights Implications of the Extraction of Lithium Perpetrated by the Tech Industry in Latin American Ecosystems. In *Technology, the Environment and a Sustainable World* (pp. 160–164). Global Information Society Watch. https://giswatch.org/node/6247

Tiku, N. (2023, July 5). *How Elite Schools Like Stanford Became Fixated on the AI Apocalypse*. The Washington Post. https://www.washingtonpost.com/technology/2023/07/05/ai-apocalypse-college-students/

Van Wynsberghe, A. (2021). Sustainable AI: AI for Sustainability and the Sustainability of AI. *AI and Ethics*, *1*(3), 213–218. https://doi.org/10.1007/s43681-021-00043-6

Vaughan, H., Pasek, A., Silcox, N. R., & Starosielski, N. (2023). *ICT Environmentalism and the Sustaianbility Game*. https://doi.org/10.1075/jlp.22125.vau

Velkova, J. (2016). Data that Warms: Waste Heat, Infrastructural Convergence and the Computation Traffic Commodity. *Big Data & Society*, *3*(2). https://doi.org/10.1177/2053951716684144

Voskoboynik, D. M., & Andreucci, D. (2022). Greening Extractivism: Environmental Discourses and Resource Governance in the 'Lithium Triangle'. *Environment and Planning E: Nature and Space*, *5*(2), 787–809. https://doi.org/10.1177/25148486211006345

Widder, D. G., & Nafus, D. (2023). Dislocated Accountabilities in the 'AI Supply Chain': Modularity and Developers' Notions of Responsibility. *Big Data & Society*, *10*(1). https://doi.org/10.1177/20539517231177620
32